# Quantum information tapping using a fiber optical parametric amplifier with noise figure improved by correlated inputs

Xueshi Guo[1], Xiaoying Li[1*], Nannan Liu[1] and Z. Y. Ou[2,+]

[1] College of Precision Instrument and Opto-electronics Engineering, Tianjin University, Key Laboratory of Optoelectronics Information Technology, Ministry of Education, Tianjin, 300072, P. R. China

[2] Department of Physics, Indiana University-Purdue University Indianapolis, Indianapolis, IN 46202, USA

Corresponding authors: *xiaoyingli@tju.edu.cn and + zou@iupui.edu

Abstract

One of the important function in optical communication system is the distribution of information encoded in an optical beam. It is not a problem to accomplish this in a classical system since classical information can be copied at will. However, challenges arise in quantum system because extra quantum noise is often added when the information content of a quantum state is distributed to various users. Here, we experimentally demonstrate a quantum information tap by using a fiber optical parametric amplifier (FOPA) with correlated inputs, whose noise is reduced by the destructive quantum interference through quantum entanglement between the signal and the idler input fields. By measuring the noise figure of the FOPA and comparing with a regular FOPA, we observe an improvement of $0.7\pm0.1$ dB and $0.84\pm0.09$ dB from the signal and idler outputs, respectively. When the low noise FOPA functions as an information splitter, the device has a total information transfer coefficient of $T_s + T_i = 1.47 \pm 0.2$, which is greater than the classical limit of 1. Moreover, this fiber based device works at the 1550 nm telecom band, so it is compatible with the current fiber-optical network.

## Introduction

In a communication network, information distribution requires information splitting. The easiest and usual way to do so is to split (tap) the signal by using a beam splitter (BS). However, in this case, the signal to noise ratio (SNR) of the tapped signal $SNR_{out}$ is lower than that of the input signal $SNR_{in}$. The reason for this degradation is well known in quantum optics. Because of the vacuum noise introduced from the unused port of BS, the sum of the information transfer coefficient in the each output port of the beam splitter, defined as $SNR_{out}/SNR_{in}$, is less than 1 (classical limit) [1]. In order to read out information without degrading the SNR, the extra noise added in the information distribution process need to be minimized. So far, it has been proposed and demonstrated that a quantum information tap with an overall information transfer efficiency larger than 1 and close to 2 in the ideal case is achievable when the vacuum noise from the unused port is squeezed [1-6].

In this paper, we demonstrate a quantum information tap by using a method that is different from the squeezed state approach. We apply the technique of quadrature amplitude entanglement for quantum noise reduction in a fiber optical parametric amplifier (FOPA) [7, 8].

In addition to demonstrating the quantum noise reduction in the amplification by exploiting the quantum destructive interference effect through quadrature amplitude entanglement of the input signals [9, 10], we investigate the noise figure (NF) of the FOPA with dual input channels. Our investigation show the NF of the FOPA with correlated inputs is better than that of a regular FOPA with one unused input port. Since the NF is inversely proportional to the corresponding information transfer coefficient, therefore, our experimental system, working at 1550 nm telecom band and compatible with the current optical fiber network, can act as the quantum information tap to split the encoded information into the signal and idler output ports with an overall information transfer coefficient larger than the classical limit.

## Experimental Principle

For a high gain phase insensitive optical parametric amplifier, there are two output ports with nearly identical powers [11-14]. Apart from amplifying the input signal, the amplifier can be used to distribute the signal input to the two output ports. When the amplifier is operated with internal mode unattended, the sum of the information transfer coefficient at two outputs is less than the classical limit of one because the corresponding NF of each output is 3 dB, which is the same as a 50/50 BS [12]. In our experiment, we overcome the 3 dB NF limit of a phase insensitive amplifier by exploiting correlated inputs. When the noise of the internal mode of the amplifier is quantum mechanically correlated to the signal input, the quantum noise at the two output ports can be partially canceled, which results in significant improvement of NF. Thus the input signal can be distributed into two outputs with reduced quantum noise, leading to quantum information tapping. In fact, the idea of using entangled quantum sources for the reduction of the extra quantum noise in a phase insensitive amplifier was proposed about two decades ago [9].

Recently, a proof in principle experiment was implemented in an atomic system by demonstrating the noise reduction at one output port [10]. Here, we will further study the NF of the amplifier with correlated inputs, which is the key parameter for evaluating the noise performance and information transfer coefficient of this kind of system. Moreover, considering the communication through optical fibers has become wide spread, we will investigate the quantum information tap by using a fiber system in 1550 nm telecom band.

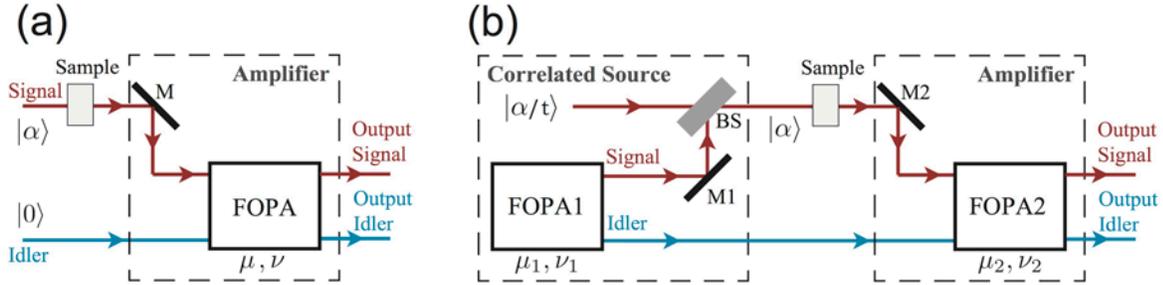

Figure 1: Conceptual diagram of the low noise amplifiers. (a) Parametric amplifier with its internal mode (idler mode) in vacuum. (b) Parametric amplifier with its internal mode (idler mode) correlated with the input signal mode. FOPA1 generates spontaneous signal and idler fields with correlated noise. The signal field is combined with a coherent state $|\alpha/t\rangle$ to produce a probe field with a non-zero coherent component $|\alpha\rangle$ and correlated with the idler field. After passing through a sample for information coding, the bright encoded probe and the spontaneous idler field are input to FOPA2 at the signal and idler input ports, respectively. In this way, the noise of the input signal is correlated with the amplifier's internal mode. Both the signal and idler outputs contain the amplified signal but with noise reduced. BS: beam splitter with near unity reflection coefficient; M1, M2: high reflection mirrors.

In our experiment, we utilize a four-wave mixing (FWM) process in optical fiber to achieve parametric amplification. Fig. 1 shows the conceptual diagram of this low noise fiber optical parametric amplifier (FOPA) using correlated source as input. In Fig. 1(a), one FOPA is shown. There are two inputs and the corresponding outputs, which are conventionally named "signal" and "idler", respectively. Normally, the idler port is not used. This is the phase insensitive amplifier used in most of the amplification applications [14]. In Fig.1 (b), there are two FOPAs in the scheme. The first one acts as a correlated quantum field generator, which, when operated without any coherent input and therefore with only vacuum input, produces two highly entangled quantum fields, dubbed also as signal and idler beams by convention [7, 8]. Because of the quantum entanglement, these two entangled quantum fields have highly correlated noise but each by themselves are in a thermal state or amplified vacuum state and do not have a non-zero coherent component [7]. For the purpose of encoding information, we will need to have a beam with non-zero coherent part propagate through a test sample. We achieve this by combining a bright coherent state $|\alpha/t\rangle$ with one (signal field) of the entangled fields generated from FOPA1 in a beam splitter (BS) with amplitude transmittance $t$. The bright encoded probe field is input into the signal input port of the second FOPA (FOPA2). This second FOPA is the amplifier whose performance we are interested in. To study the amplifier with correlated input, we will inject the other one (idler) of the entangled fields from FOPA1 into the idler input port of FOPA2. Then the information encoded from the sample is amplified by FOPA2 and comes out at both the signal and idler output ports of the amplifier.

When the amplitude reflectance of the BS satisfies $r \approx 1$ (the amplitude transmittance $t$ is related with $r$ by $t^2 + r^2 = 1$), the noise level of the signal and idler inputs is governed by the correlated fields from FOPA1. But the signal level for the input of FOPA2, which carries the information of the sample, has a non-zero coherent part of $|\alpha\rangle$.

Next, we first briefly analyze the quantum noise and the SNR for the amplification scheme shown in Fig. 1(b) and compare it with the scheme in Fig. 1(a). Assuming a signal mode (denoted by $\hat{a}_s$) and an idler mode (denoted by $\hat{a}_i$) are sent to an amplifier based on FWM, the operators of the signal and idler output fields, $\hat{b}_s$ and $\hat{b}_i$ can be written as [15, 16]:

$$\hat{b}_{s(i)} = \mu \hat{a}_{s(i)} + e^{2i\theta_p} \nu \hat{a}^\dagger_{i(s)}, \tag{1}$$

where $\mu = \cosh g$ and $\nu = \sinh g$ with $g > 0$ denoting the nonlinear coupling coefficient of FWM are referred to as the amplitude gain of the amplifier, and $\theta_p$ is the phase of the pump field. The experimentally measurable power gain of the phase insensitive amplifier is $\mu^2$.

For a given mode $\hat{a}$, the quadrature-phase amplitude operator, defined as $\hat{X}(\theta) = e^{-i\theta}\hat{a} + e^{i\theta}\hat{a}^\dagger$, is more often used to convey information since it can be directly detected by a homodyne detector. The quadrature amplitude operator $\hat{X}_k$ can be written as the summation of its mean value $\langle \hat{X}_k \rangle$ and noise fluctuation $\Delta \hat{X}_k$:

$$\hat{X}_k = \langle \hat{X}_k \rangle + \Delta \hat{X}_k, \text{ for } k = s0, i0, s, i \tag{2}$$

where $s0, i0$ represent the signal and idler input channels, and $s, i$ represent the signal and idler output channels, respectively. So, the mean value of signal input is $\langle \hat{X}_{s0}(\theta) \rangle = \alpha e^{-i\theta} + \alpha^* e^{i\theta}$. When the phase of two pumps for FOPA1 and FOPA2 in Fig. 1(b) are set to be $\theta_{p1}$ and $\theta_{p2}$, the stimulated part of the signal and idler outputs, which determines the mean value of the quadrature operator and conveys the information of the sample, can be obtained by using Eq. (1):

$$\langle \hat{X}_s(\theta) \rangle = \mu_2 \left( \alpha e^{-i\theta} + \alpha^* e^{i\theta} \right), \tag{3}$$

$$\langle \hat{X}_i(\theta) \rangle = v_2 \left[ \alpha e^{-i(\theta + 2\theta_{p2})} + \alpha^* e^{i(\theta + 2\theta_{p2})} \right], \quad (4)$$

where $\mu_2$ and $v_2$ are amplitude gain of FOPA2. Note that the phase angle for the signal output field remains the same as the coherent state $|\alpha\rangle$, but the phase angle of the idler output field is shifted by $2\theta_{p2}$ because of the FWM process.

The noise level for the input of this amplifier is determined by the spontaneous fields from FOPA1. Therefore, the coherent signal input can be ignored when we analyze the noise performance of the amplifier. Using Eq. (1) twice for the scheme in Fig. 1(b), the operator of the quadrature amplitude noise for signal (idler) output $\Delta \hat{X}_{s(i)}(\theta)$ can be written as:

$$\Delta \hat{X}_{s(i)}(\theta) = \mu_1 \mu_2 \Delta \hat{X}_{s0(i0)}(\theta) + v_1 v_2 \Delta \hat{X}_{s0(i0)}(\theta + 2(\theta_{p1} - \theta_{p2})) \\ + v_1 \mu_2 \Delta \hat{X}_{i0(s0)}(-\theta + 2\theta_{p1}) + \mu_1 v_2 \Delta \hat{X}_{i0(s0)}(-\theta + 2\theta_{p2}), \quad (5)$$

where $\mu_1$, $v_1$ and $\mu_2$, $v_2$ are the amplitude gain for FOPA1 and FOPA2, respectively, $\hat{X}_{s0(i0)}$ with noise variance $\langle (\Delta \hat{X}_{s0(i0)})^2 \rangle = 1$ denotes the quadrature amplitude operator of vacuum at the signal (idler) input port of FOPA1. Then the output noise level for arbitrary quadrature angle $\theta$ can be found as

$$\langle (\Delta \hat{X}_s(\theta))^2 \rangle = \langle (\Delta \hat{X}_i(\theta))^2 \rangle = \mu_1^2 \mu_2^2 + v_1^2 v_2^2 + \mu_1^2 v_2^2 + v_1^2 \mu_2^2 + 4\mu_1 \mu_2 v_1 v_2 \cos[2(\theta_{p2} - \theta_{p1})]. \quad (6)$$

From Eq. (3), one sees that the average of the signal field amplified by FOPA2 does not depend on the phases of the pumps indicating a phase insensitive amplifier, but Eq. (6) shows that the noise of the output changes with the phases of the pumps. When $\theta_{p2} - \theta_{p1} = \pi/2$, the output noise is minimized and can be written as:

$$\langle(\Delta \hat{X}_s(\theta))^2\rangle = \langle(\Delta \hat{X}_i(\theta))^2\rangle = \mu_1^2\mu_2^2 + \nu_1^2\nu_2^2 + \mu_1^2\nu_2^2 + \nu_1^2\mu_2^2 - 4\mu_1\mu_2\nu_1\nu_2 \\ = 1 + 2(\mu_1\nu_2 - \mu_2\nu_1)^2, \quad (7)$$

in which the relations of the amplitude gain $\mu_1^2 - \nu_1^2 = 1$ and $\mu_2^2 - \nu_2^2 = 1$ are applied. So, the SNRs for the signal and idler outputs of FOPA2 are

$$SNR_s = \frac{\langle \hat{X}_s(\theta)\rangle^2}{\langle(\Delta \hat{X}_s(\theta))^2\rangle} = \frac{\mu_2^2\left(\alpha e^{-i\theta} + \alpha^* e^{i\theta}\right)^2}{1 + 2(\mu_1\nu_2 - \mu_2\nu_1)^2}, \quad (8)$$

$$SNR_i = \frac{\langle \hat{X}_i(\theta)\rangle^2}{\langle(\Delta \hat{X}_i(\theta))^2\rangle} = \frac{\nu_2^2\left[\alpha e^{-i(\theta+2\theta_{p2})} + \alpha^* e^{i(\theta+2\theta_{p2})}\right]^2}{1 + 2(\mu_1\nu_2 - \mu_2\nu_1)^2}. \quad (9)$$

Note that without the correlated quantum fields, i.e., $\mu_1 = 1$, $\nu_1 = 0$, which corresponds to the scheme in Fig. 1(a), the output SNRs of the amplifier will be

$$SNR_{s'} = \frac{\mu_2^2\left(\alpha e^{-i\theta} + \alpha^* e^{i\theta}\right)^2}{\mu_2^2 + \nu_2^2} \quad (10)$$

and

$$SNR_{i'} = \frac{\nu_2^2\left(\alpha e^{-i\theta} + \alpha^* e^{i\theta}\right)^2}{\mu_2^2 + \nu_2^2} \quad (11)$$

in the signal and idler output ports, so the improvement in SNR due to the application of the correlated input fields is

$$\frac{SNR_s}{SNR_{s'}} = \frac{SNR_i}{SNR_{i'}} = \frac{\mu_2^2 + v_2^2}{1 + 2(\mu_1 v_2 - \mu_2 v_1)^2}. \tag{12}$$

This value reaches the maximum of $\mu_2^2 + v_2^2$ when $\mu_1 = \mu_2, v_1 = v_2$. In this case, FOPA2 is a reverse process of FOPA1. According to Eq. (7), the output noise of the amplifier is simply $\langle(\Delta \hat{X}_s(\theta))^2\rangle = 1$, which means that the output noise level of FOPA2 remains the same as the vacuum state even with amplification. It is reduced by a factor of $1/(\mu_2^2 + v_2^2)$ from the case without correlated inputs. This is because of a destructive quantum interference effect for noise cancelation [10]. Hence, we have the improvement in SNR as $SNR_s/SNR_{s'} = \mu_2^2 + v_2^2$.

Since the noise performance of an amplifier is often characterized by NF, defined as the ratio of input SNR to the output SNR, let us now compare the NFs for the schemes with correlated sources in Fig. 1(b) and a regular amplifier in Fig. 1(a). To do so, we need to know the SNR at the input port. For the regular amplifier with the coherent state input (Fig. 1(a)), it is simply $SNR_{in'} = (\alpha e^{-i\theta} + \alpha^* e^{i\theta})^2$ because the noise is the shot noise or vacuum noise of 1. So the noise figures for the signal and idler output ports are written as

$$NF_s' \equiv \frac{SNR_{in'}}{SNR_{s'}} = \frac{v_2^2}{\mu_2^2} + 1 \tag{13}$$

and

$$NF_i' \equiv \frac{SNR_{in'}}{SNR_{i'}} = \frac{\mu_2^2}{v_2^2} + 1. \tag{14}$$

Notice that $NF_s' \to 2$ for $\mu_2 \to \infty$, leading to the famous 3 dB value in a regular amplifier. For the scheme with correlated inputs in Fig. 1(b), the average value of input signal is still $\langle \hat{X}_{s0}(\theta) \rangle = \alpha e^{-i\theta} + \alpha^* e^{i\theta}$, however, the idler input port is now occupied by the idler output of FOPA1. Therefore, the noise at the input port is determined by both the inputs in signal and idler channels. Since the quadrature-amplitude noise at the signal and idler inputs of FOPA2, respectively denoted by the operators $\Delta \hat{X}'_{s0}(\theta)$ and $\Delta \hat{X}'_{i0}(\theta)$, are correlated, the effective noise for the dual channel input case is then expressed as

$$\Delta \hat{X}_c(\theta) = \Delta \hat{X}'_{s0}(\theta) - \lambda \Delta \hat{X}'_{i0}(\theta) \\ = [\mu_1 \Delta X_{s0}(\theta) + \nu_1 \Delta X_{i0}(2\theta_{p1} - \theta)] - \lambda [\mu_1 \Delta X_{i0}(\theta) + \nu_1 \Delta X_{s0}(2\theta_{p1} - \theta)], \quad (15)$$

where the coefficient $\lambda$ can be adjusted by changing the electronic gain of the detection system. It is straightforward to deduce that the minimum value of the effective noise is $\langle (\Delta \hat{X}_c(\theta))^2 \rangle = \frac{1}{\mu_1^2 + \nu_1^2}$ under the conditions of $\lambda = 2\nu_1 \mu_1 / (\mu_1^2 + \nu_1^2)$ and $\theta_{p1} = \theta$. Thus, the input SNR in the dual channel case is $SNR_{in} = (\alpha e^{-i\theta} + \alpha^* e^{i\theta})^2 (\mu_1^2 + \nu_1^2)$. Consequently, the noise figures for the two outputs in Fig. 1(b) are

$$NF_s \equiv \frac{SNR_{in}}{SNR_s} = \left(1 + 2(\mu_1 \nu_2 - \mu_2 \nu_1)^2\right) \frac{\mu_1^2 + \nu_1^2}{\mu_2^2}, \quad (16)$$

$$NF_i \equiv \frac{SNR_{in}}{SNR_i} = \left(1 + 2(\mu_1 \nu_2 - \mu_2 \nu_1)^2\right) \frac{\mu_1^2 + \nu_1^2}{\nu_2^2}. \quad (17)$$

When $\mu_2 = \mu_1^2 + \nu_1^2$, $NF_s$ has the minimum value of 1 and $NF_i = \mu_2^2 / \nu_2^2$. Notice that the optimum condition here ($\mu_2 = \mu_1^2 + \nu_1^2$) is different from the minimum output noise condition ($\mu_2 = \mu_1$) discussed in Eq. (12) because we need to consider the SNR at input port, which is different in the two schemes in Fig. 1. On the other hand, at the optimum condition of $\mu_2 = \mu_1^2 + v_1^2$, we have $NF_s = 1$, which corresponds to noiseless amplification. Comparing this with Eqs. (13) and (14), the improvement in noise figure is expressed as a reduction of

$$\frac{NF_s}{NF_s'} = \frac{NF_i}{NF_i'} = \frac{\mu_2^2}{\mu_2^2 + v_2^2} < 1. \tag{18}$$

Hence, the noise figure is reduced for both outputs of the amplifier when we use correlated sources in the amplification process.

According to the noise figure of the scheme in Fig. (b), it is straightforward to deduce the information transfer coefficients for the signal and idler output ports:

$$T_s \equiv \frac{SNR_s}{SNR_{in}} = \frac{1}{NF_s}, \quad T_i \equiv \frac{SNR_i}{SNR_{in}} = \frac{1}{NF_i}. \tag{19}$$

According to Eqs. (16) and (17), it is obvious that the maximum values for $T_s$ and $T_i$ occur when the NF of the amplifier is optimized, i.e.,

$$T_s + T_i = 1 + v_2^2/\mu_2^2 \to 2 \quad \text{for} \quad \mu_2 = \mu_1^2 + v_1^2 \text{ and } \mu_2 \to \infty. \tag{20}$$

Notice that the overall transfer coefficient $T_s + T_i$ is larger than the classical limit 1 [1]. Thus, such a device can function as a quantum information tap to split an input signal into two copies without adding noise. It is worth noting that the quantum clone amplifier in Ref. [2] also has two amplified copies of the input. But here, we achieve the same with a phase-insensitive amplifier instead of a phase-sensitive one.

## Results

**Experimental setup**

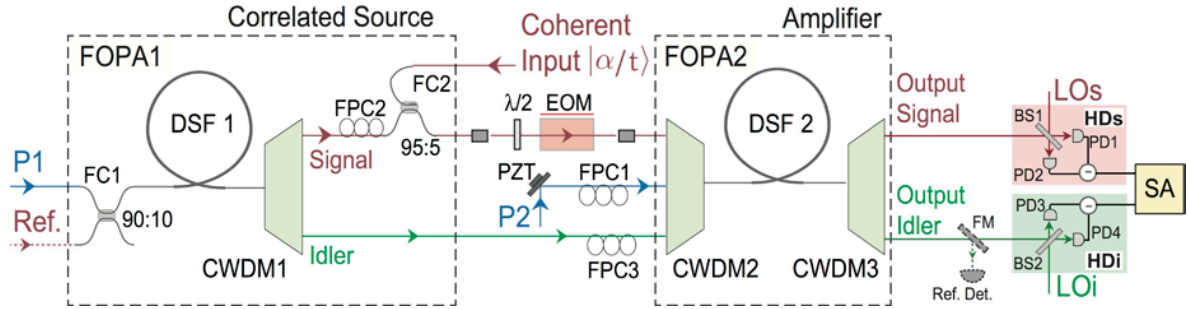

Figure 2: Experimental setup of the low noise amplifier with correlated noise inputs. P1, P2, pulsed pumps; DSF, dispersion shifted fiber; CWDM, coarse wavelength division multiplexer; FPC, fiber polarization controller; FC, fiber coupler; EOM, electro-optics modulator; LOs, LOi, local oscillators; HD, homodyne detection; SA, electronic spectrum analyzer; FM, flip mirror; Ref. reference light; Ref. Det., reference detector.

The experimental setup for realizing the scheme in Fig. 1(b) is shown in Fig. 2. The nonlinear media for FOPA1 and FOPA2 are two pieces of dispersion shifted fiber (DSF1 and DSF2) with identical dispersion properties. The length and zero dispersion wavelength of each DSF are about 1550 nm and 300 m, respectively. The pumps P1 and P2 for DSF1 and DSF2 are the same and are centered at 1552.5 nm. Since the wavelength of the pumps is in the anomalous dispersion region of DSF1/DSF2, the phase matching condition of FWM with broad gain bandwidth is satisfied. In DSF1, the pulsed pump P1 generates the correlated signal (1569 nm) and idler (1534 nm) fields through spontaneous FWM [17]. After being separated with a standard coarse wavelength division multiplexer (CWDM1), the spontaneous signal field is combined with a pulsed coherent state input signal at a 95:5 fiber coupler (FC2). The combined signal beam together with the separated spontaneous idler field from DSF1 forms a correlated quantum source with a non-zero coherent component. To produce an input test signal for the amplifier, the combined signal beam is encoded by phase modulation (PM) with an electro-optic modulator (EOM). The encoded signal beam and the correlated idler beam are combined with the pump P2 by CWDM2 and are sent to DSF2 to realize the low noise amplification. At the output of DSF2, the amplified fields and the residual pump field are separated by CWDM3. Since the noise level of FOPA2 is sensitive to the phase difference of the pumps P1 and P2 (see Eq. (6)), we control the phase of P2 by mounting a piezo-electric transducer (PZT) on a high reflection mirror. In the experiment, all the CWDMs have four channels centered at 1571 nm (for the signal), 1551 nm (for the pump), 1531 nm (for the idler) and 1511 nm (not used), respectively, and the 1-dB bandwidth of each channel is about 16 nm. The quadrature-phase amplitudes of the signal and idler outputs of FOPA2 are detected with two homodyne detectors HDs and HDi, whose local oscillators are labelled as LOs' and LOi (see Fig. 2), respectively. The performance of FOPA2 is investigated by analyzing the photo-currents of HDs and HDi with an electronic spectrum analyzer (SA).

**Noise Reduction**

We first investigate the quantum noise of FOPA2. In the measurement, the coherent state input $|\alpha/t\rangle$ is blocked. The powers of P1 and P2 are 1.1 and 1.7 mW, respectively, and the corresponding power gain for FOPA1 and FOPA2 are about 6 and 20, respectively. Figs. 3(a) and 3(b) show the noise measured by HDs and HDi at the signal and idler channels, respectively. The shot noise levels (SNLs) of the homodyne detection systems (brown traces) are obtained by blocking the pumps P1 and P2 to ensure that only vacuum state is at the output ports. The plots in the left column of Fig. 3 present the results for FOPA2 with the correlated inputs. The green traces are obtained by scanning the phase of P2 while red traces are obtained by setting the phase of P2 to achieve the lowest noise level. These results agree with Eq. (6): the output noise level varies with the relative phase difference between P1 and P2.

As a comparison, we also measure the noise when FOPA2 functions as a regular amplifier. The plots in the middle of Fig. 3 are obtained by blocking the idler output of FOPA1. In this situation, there is thermal noise in the signal input port. In addition to measuring the noise of the output ports (black traces), we record the noise of input ports (pink traces) by blocking P2 to ensure DSF2 is simply a regular transmission fiber. Clearly, the level of pink trace in Fig. 3(b), representing the noise of the unused idler input port, is the same as the SNL of HDi (brown trace); while the pink trace in Fig. 3(a) is higher than the SNL of HDs (brown trace) because of the thermal nature of the individual signal field generated by FWM in DSF1. Comparing with the noise for FOPA2 with correlated inputs, we find that the black traces are higher than the red traces by about 2.1 dB and 2.5 dB in Fig. 3(a) and Fig. 3(b), respectively. This result indicates that the correlation between the signal and the idler fields plays an important role in reducing the output noise of FOPA2.

The blue traces in the right plots of Fig. 3 are the noise at the output ports of a regular FOPA with shot noise at the input port, which is obtained by blocking P1. We find that the minimum output noise level of FOPA2 with correlated inputs (red trace) is even lower than blue trace by about 1 dB in both Figs. 3(a) and 3(b). Therefore, similar to squeezed states, amplifiers with quantum correlated sources as inputs can also outperform the amplifier with classical sources as inputs. The mechanism for the noise reduction is destructive quantum interference through quantum entanglement between the signal and the idler fields.

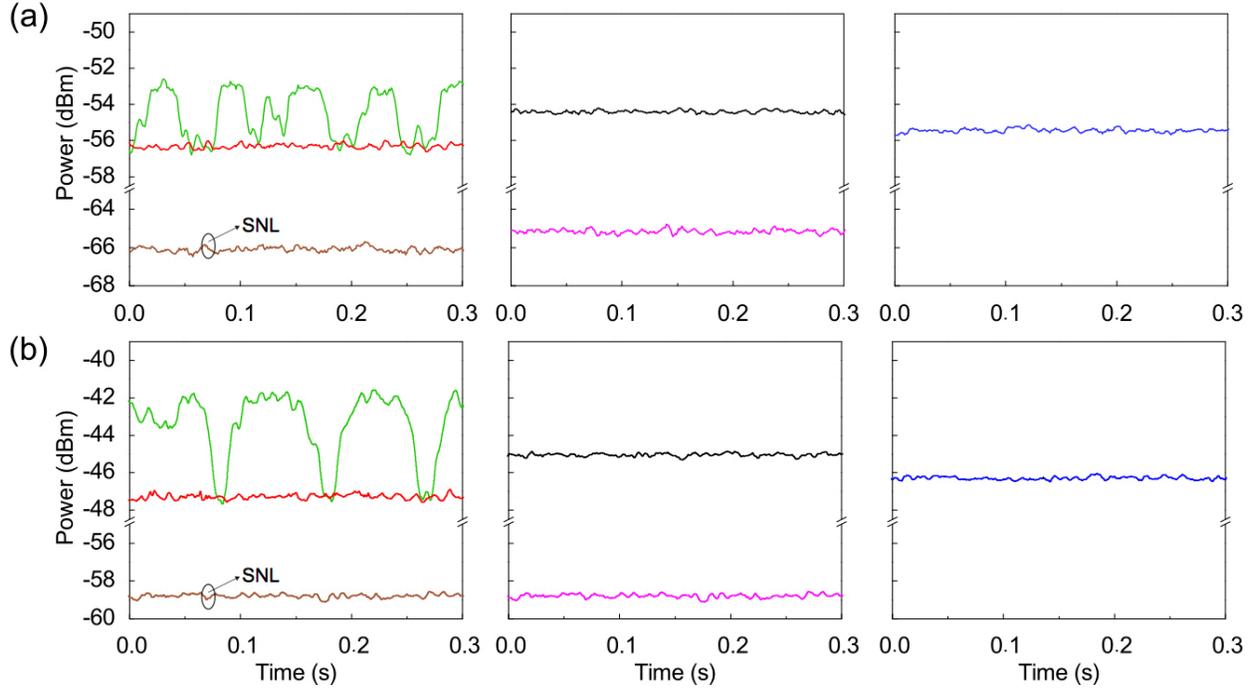

Figure 3: The noise levels measured at (a) the signal and (b) the idler output ports of FOPA2 under different conditions. The plots in the left column are obtained for FOPA2 with correlated inputs, while the plots in the middle and right columns are achieved when FOPA2 functions a regular FOPA with thermal noise input and shot noise input, respectively. The brown traces are the shot noise levels (SNL) of the homodyne detection systems; the green and red traces are obtained by scanning the phase of P2 and setting the phase of P2 to achieve the lowest noise levels, respectively; the black and pink traces are respectively measured by blocking the idler output field of DSF1 and by blocking both the idler output field of DSF1and P2; blue traces are obtained by blocking P1. In the measurement, SA is set to zero span at 2 MHz and the electronic noise of the homodyne detection system is subtracted.

## Low noise amplification

We then test FOPA2 with a phase modulated (PM) weak signal and measure the NF to demonstrate its application in quantum information processing. During the test, a sinusoidal modulation voltage at frequency of 2 MHz is applied on the EOM, the powers of the pumps P1 and P2 are still fixed at 1.1 and 1.7 mW, respectively, and the power of the weak coherent input light propagating through the EOM is 2 nW. The information encoded through the EOM is amplified by FOPA2, and the NF is analyzed by measuring the SNR at the input and the output ports. We first analyze the NF of the FOPA2 with correlated inputs when the phase of P2 is fixed at the point to achieve the lowest noise. As a comparison, we then analyze the NF when FOPA2 functions as the regular amplifier with (i) thermal noise input and (ii) shot noise input, respectively. The results are shown in Fig. 4, in which the red traces correspond to the measurement for FOPA2 with correlated inputs, while the black and blue traces correspond to the measurement for the regular FOPA in cases (i) and (ii), respectively. For convenience, we summarize and list in Table 1 the SNRs and NFs measured from the curves in Fig. 4.

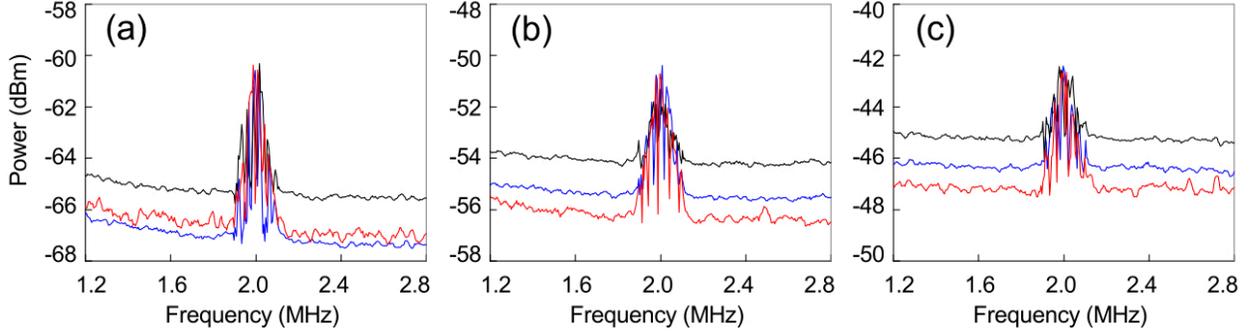

Figure 4: (a) The phase modulated (PM) signal and noise levels of the input signal port of FOPA2. (b) and (c) are the PM signal and noise levels of the signal and the idler output ports of FOPA2, respectively. Blue traces, obtained by blocking the pump P1, correspond to the case of a regular amplifier with shot noise input; black traces, obtained by blocking the idler input port of FOPA2, correspond to the case of the a regular amplifier with thermal noise input; and red traces, obtained by launching the correlated signal and idler fields into FOPA2, correspond to the case of the amplifier with correlated inputs.

Figure 4(a) shows the measured phase modulated (PM) signal and noise levels of the signal input port of FOPA2 in different cases. During the measurement, pump P2 is blocked so that FOPA2 is a regular passive fiber ($\mu_2 = 1$, $\nu_2 = 0$). We can see that the peaks of the modulated signal in three traces are almost the same, but their noise levels are different. The noise level at 2

MHz of each trace is obtained by linear extrapolation of the noise level in the frequency range of 1.2-1.9 MHz and 2.1-2.8 MHz. For the regular FOPA with shot noise input (blue trace), the noise level is equivalent to the SNL (brown trace) in Fig. 3(a), and the input SNR is $SNR_{in'} = 6.69 \pm 0.05$ dB. For the regular FOPA with thermal noise input (black trace), the noise level is equivalent to the pink trace in Fig. 3(a) and is about 2 dB higher than that of blue trace because the individual signal field out of FOPA1 is of thermal nature. So the input SNR (black trace) degrades to $SNR_{in''} = 4.93 \pm 0.04$ dB. While for FOPA2 with dual channel inputs, the result (red trace) is obtained by subtracting the photo-current of HDs from that of HDi. One sees that the noise level of the photo-current difference is about 1.5 dB lower than that of black trace due to the quantum noise correlation between the signal and idler fields from FOPA1 [8]. Comparing the black and red traces, it is clear that the presence of the correlated idler field improves the SNR from $SNR_{in''} = 4.93 \pm 0.04$ dB to $SNR_{in} = 6.49 \pm 0.05$ dB. We note that according to the theoretical analysis (see Eq. (15)), the noise of the red trace should be lower than that of blue trace due to Einstain-Podosky-Rosen type quantum noise correlation [7, 8], however, in our experiment, the noise level of the red trace is slightly higher than that of blue trace. We believe this is because of the relatively low transmission efficiency (68%) between DSF1 and DSF2 [17, 18].

Figure 4(b) [4(c)] demonstrates the amplified PM signal and noise level at the signal [idler] output port of FOPA2. For the regular FOPA with shot noise input (blue trace), its noise is equivalent to the blue trace in Fig. 3(a) [3(b)], the output SNR is $SNR_{s'} = 5.14 \pm 0.05$ [ $SNR_{i'} = 3.94 \pm 0.04$ ] dB, which indicates that the NF is $NF_{s}^{'} \equiv SNR_{in'}/SNR_{s'} = 1.55 \pm 0.07$ [ $NF_{i}^{'} = 2.75 \pm 0.07$ ] dB at the signal [idler] port. For the regular FOPA with thermal noise input

(black trace), its noise level is equivalent to the black trace in Fig. 3(a) [3(b)], the output SNR is $SNR_{s''} = 3.42 \pm 0.05$ [ $SNR_{i''} = 2.23 \pm 0.07$ ] dB, and the NF is $NF_s'' = 1.51 \pm 0.08$ [ $NF_i'' = 2.70 \pm 0.07$ ] dB at the signal [idler] port, which is about the same as that of the regular amplifier with shot noise input (blue trace). For FOPA2 with correlated inputs (red trace), its noise level equivalents to the red trace in Fig. 3(a) [3(b)], the output SNR is $SNR_s = 5.64 \pm 0.05$ [ $SNR_i = 4.58 \pm 0.05$ ] dB, leading to a NF of $NF_s = 0.85 \pm 0.08$ [ $NF_i = 1.91 \pm 0.07$ ] dB, which is better than that of the regular FOPA.

Comparing the red and blue traces in Fig. 4, we find that the improvement of the NF in signal [idler] channel due to the noise cancelation of the correlated inputs is $NF_s'/NF_s = 0.70 \pm 0.10$ [ $NF_i'/NF_i = 0.84 \pm 0.09$ ] dB. Moreover, for the FOPA2 with correlated inputs, we deduce the information transfer efficiencies of signal and idler (tap) port $T_{s,i}$, and find

$$T_s = \frac{SNR_s}{SNR_{in}} = 5.64 - 6.49 = -0.85 dB = 0.83 \quad \text{and} \quad T_i = \frac{SNR_i}{SNR_{in}} = 4.58 - 6.49 = -1.91 dB = 0.64 \quad .$$

Hence, we have $T_s + T_i = 1.47 \pm 0.2 > 1$. Since the limit for any classical information tap is 1, we have realized a quantum information tap in the two outputs of the amplifier coupled to the correlated sources.

|  | Regular FOPA | | FOPA2 with correlated inputs |
|---|---|---|---|
|  | With shot noise input | With thermal noise input |  |
| SNR | $SNR_{in'} = 6.69 \pm 0.05$ dB | $SNR_{in''} = 4.93 \pm 0.04$ dB | $SNR_{in} = 6.49 \pm 0.05$ dB |
|  | $SNR_{s'} = 5.14 \pm 0.05$ dB | $SNR_{s''} = 3.42 \pm 0.05$ dB | $SNR_s = 5.64 \pm 0.05$ dB |
|  | $SNR_{i'} = 3.94 \pm 0.04$ dB | $SNR_{i''} = 2.23 \pm 0.07$ dB | $SNR_i = 4.58 \pm 0.05$ dB |
| NF | $NF_s' = 1.55 \pm 0.07$ dB | $NF_s'' = 1.51 \pm 0.08$ dB | $NF_s = 0.85 \pm 0.08$ dB |

|  | $NF_i^{'} = 2.75 \pm 0.07$ dB | $NF_i^{''} = 2.70 \pm 0.07$ dB | $NF_i = 1.91 \pm 0.07$ dB |
|---|---|---|---|

Table1: The signal to noise ratio (SNR) and the noise figure (NF) of FOPA in different cases.

## Discussion

Figs. 3(a) and 3(b) are very similar, but there are two obvious differences between them. First, under the same experimental condition, the measured noise level in Fig. 3(b) is higher than that in Fig. 3(a). This is because the power of LOi (2 mW) is about 5 times higher than that of LOs. Second, the difference between the maximum and the minimum values of the green trace in Fig. 3(b) is larger than that in Fig. 3(a). We believe this is due to the difference between the mode matching efficiencies of HDs and HDi. In the experiment, the total detection efficiencies at both the signal and idler output ports are 48% when the transmission efficiencies of about 70% between the output end of DSF1 and the input end of DSF2 and 80% between the output end of DSF2 and the input end of HDs/HDi, and the detection efficiency of about 85% for both HDs and HDi are included. But the temporal mode matching efficiencies of HDs and HDi, which is estimated by injecting the coherent input $|\alpha/t\rangle$, while blocking P1, and measuring the visibility of interference between the LOs/LOi and the amplified signal/idler, are about 72% and 84%, respectively. Therefore, for the result of green traces, improving the temporal mode matching of HDs/HDi might increase the difference between the measured maximum and minimum noise levels.

Moreover, it is worth noting that the SNR can be as high as needed if one can use intense enough light to carry information, however, there are obvious practical limitations to the energy of the optical field. For example, for the applications of quantum information processing and biological sensing, the intensity of optical beam used to encode information is usually very weak.

With the decrease of the intensity of the optical field, the corresponding SNR will be quickly limited by the standard quantum limit. In our experiment, the power propagating through the sample (EOM) is in the nanowatt level or even weaker, which may too weak to exceed the electronic noise of a photodiode detector. After passing the weak signal through FOPA2 with correlated inputs, not only is the NF improved, but the power of the signal is also significantly amplified, which will help to realize the direct detection of the signal because the photocurrent of the amplified signal can surpass the electronic noise of the detector. Therefore, in addition to functioning as the quantum information tap, the FOPA2 with improved NF can be used for pre-amplifying the information carried by weak optical beams.

Furthermore, we compare the performance of our low noise phase insensitive FOPA with that of the low noise phase sensitive FOPA in Ref. [19]. We find that for the former, the encoded signal only exists in one input port, the improvement in SNR and NF depends on the noise correlation between two inputs. The output SNR is $\mu_2^2 + \nu_2^2$ times better than a regular FOPA, and NF=1 (noiseless amplification) is in principle achievable. While for the latter, the noise of the two inputs is uncorrelated, the improvement in SNR and NF depends on correlation of signal in two input ports. The output SNR is 4 times better than a regular FOPA, and the NF is at most two times better than a regular FOPA [20]. Hence, in contrast to low noise FOPA in Ref. [19], which is suitable for amplifying the classical signal with extremely high loss, the FOPA with correlated inputs, which has a greater potential on improving the SNR and NF, is suitable for amplifying the quantum signal with low loss.

## Conclusions

We have built a low noise fiber optical parametric amplifier (FOPA) with correlated input. Comparing with a regular FOPA with vacuum in the unused idler input port, the noise level of the low noise FOPA is reduced by about 1 dB in both the signal and the idler output channels. We have also analyzed and measured the noise figures for the FOPA with correlated input and compared with a regular FOPA. An improvement of $0.70 \pm 0.10$ and $0.84 \pm 0.09$ dB is observed from the signal and idler outputs, respectively. When the low noise FOPA functions as an information splitter, the device has a total information transfer efficiency of $T_s + T_i = 1.47 \pm 0.2$, which is greater than the classical limit of one. We therefore have achieved a quantum information tap. Moreover, we would like to point out that the experimental results are currently not as good as the theoretical predictions. We think this is because the noise reduction in our system is limited by the optical losses between two FOPAs, by the imperfect temporal mode-matching, and by the excess noise of the fiber laser. We believe the performance of the low noise FOPA can be further enhanced by technically taking care of the experimental imperfections and by developing a multi-mode theory for better describing the experimental system and optimizing the experimental parameters.

# Methods

## Preparation of relevant light fields

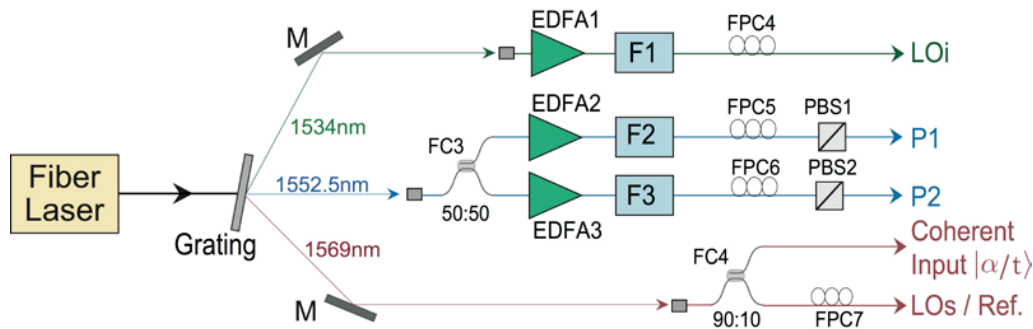

Figure 5: The scheme for the preparation of the pulsed pumps (P1, P2), the coherent state input signal, the local oscillators (LOs, LOi) and the reference signal (Ref.). FPC, fiber polarization controller; FC, fiber coupler; EDFA, erbium-doped fiber amplifier; F, filter; PBS, polarization beam splitters.

In the experiment, for the purpose of testing and alignment of the system, a reference signal (Ref.) is injected through a 90:10 fiber coupler (FC1). The preparation of the pulsed pumps (P1, P2), the coherent state input signal, the local oscillators (LOs, LOi) and the reference signal (Ref.) is shown in Fig. 5. We first disperse the 40 MHz train of 150-fs pulses centered at 1560 nm from a mode-locked fiber laser with a grating and spectrally filter them to obtain three beams, whose wavelengths are selected at 1552.5, 1569 and 1534 nm for the pumps, the signal and the idler fields, respectively. At these wavelengths, the FWM process in DSFs is most effective. We then propagate the beam at 1552.5 nm through the 50:50 fiber coupler (FC3) and amplify each output of FC3 by EDFA2 and EDFA3, respectively, to achieve the required power for the two pumps. The output of EDFA2/EDFA3 is further cleaned up with a bandpass filter F2/F3 centered at 1552.5 nm and having a FWHM of 0.8 nm. To ensure the polarization and power adjustment of the pump P1/P2, the output of F2/F3 then passes through a fiber polarization controller FPC5/FPC6 and polarization beam splitter PBS1/PBS2. LOi is achieved by amplifying the beam at 1534 nm (a much weaker spectral component from the femto-second laser) with EDFA1 and propagating through filter F1 (centered at 1534 nm with a FWHM of 1 nm). LOs is directly obtained by passing the beam at 1569 nm, with a power of about 450 μW, through the 90% port of FC4, the reference light obtained by attenuating the LOs to about 2 μW, and the coherent input $|\alpha/t\rangle$ is obtained at the 10% port of FC4.

**Realization of mode matching**

The noise reduction effect only occurs when the mode-matching is well implemented in the aspects of polarization state, spatial distribution and temporal mode for all the optical fields involved in the FWM in FOPA2. Since the spatial mode matching is automatically fulfilled in a single-mode fiber and the polarization state can be controlled by FPCs and PBSs (see Fig. 2 and Fig. 5), the key work during the experiment is to synchronize different light pulses before entering DSF2.

The mode matching is realized by the following two procedures. Firstly, we match the modes of coherent input and pump P2 in FOPA2. To do so, we block the spontaneous FWM fields from FOPA1, and observe the parametric gain of FOPA2 by adjusting the relative optical delay between the input $|\alpha/t\rangle$ and P2 for pulse overlap and by tuning the polarization of P2 with FPC1. The best mode matching is achieved when the parametric gain is maximum. Secondly, we match the modes of the spontaneous FWM fields from DSF1 with P2 of FOPA2. To do so, we block the coherent input $|\alpha/t\rangle$ and inject the reference beam (Ref.) into DSF1 for alignment. The mode match between the reference beam and P1 for DSF1 is done in the same way as the first step by adjusting their optical paths to maximize the pulse overlap and by tuning the polarization with FPC7. With the injection of reference beam, the setup in Fig. 2 is configured as a nonlinear interferometer [21]. We can observe the nonlinear interference fringe at the output idler port of CWMD3 with a flip mirror (FM) and a reference detector (Ref. Det.) by scanning the phase of P2 with the PZT. For the best mode match, we maximize the fringe visibility by carefully adjusting the relative delays and fine tuning the polarizations (with FPC2 and FPC3) of the signal, the idler (both from DSF1) and P2. In this process, the powers of P1 and P2 are about 1.1 and 1.7 mW, respectively. Fig. 6 shows the simultaneously recorded output (blue) of the reference detector and the voltage of PZT (pink). An interference fringe (blue) with nearly 100% visibility is observed, indicating that the required mode match of all the relevant fields in FOPA2 is achieved. Note that the reference light is only used for mode matching because the modes of spontaneous FWM fields follow that of the stimulated FWM fields in DSF1, it is blocked in the experiments of noise reduction and low noise amplification.

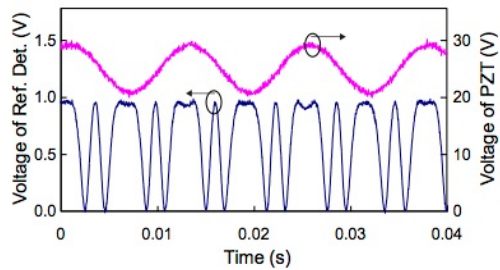

Figure 6: The nonlinear interference fringe (blue) and voltage applied on the piezo-electric transducer (pink). The fringe is obtained with a slow opto-electronic detector (Ref. Det.) at the idler output port of FOPA2 when the reference signal is injected into DSF1

The quadrature amplitudes at the signal and idler output ports of FOPA2 are measured by HDs and HDi. The homodyne detector HDs/HDi is comprised of a 50:50 BS1/BS2 and two photodiodes (PD, ETX-500) with circuits the same as those in Ref. [22]. During the measurement process, the required mode matching between the measured signal/idler output and LOs/LOi is achieved by adjusting the polarization of LOs/LOi with FPC7/FPC4 (see Fig. 2 and Fig. 5).

## Acknowledgements

This work was supported in part by the State Key Development Program for Basic Research of China (No. 2014CB340103), the National Natural Science Foundation of China (No. 11527808) , the Specialized Research Fund for the Doctoral Program of Higher Education of China (No. 20120032110055), and 111 Project B07014.

## Author contribution statement

## Competing interests

The authors declare no competing financial interests.